\documentclass[10pt,a4paper,twocolumn]{article}
\usepackage{titlesec}
\usepackage{color}
\usepackage[utf8]{inputenc} 
\usepackage[affil-it]{authblk}

\usepackage{lipsum}
\newcommand\blfootnote[1]{%
  \begingroup
  \renewcommand\thefootnote{}\footnote{#1}%
  \addtocounter{footnote}{-1}%
  \endgroup
  }
\usepackage{siunitx}
\usepackage{float}
\usepackage{tikz}
\usepackage{xcolor}
\usepackage{comment}

\usepackage{braket}
\usepackage{nicefrac}
\usepackage[english]{babel} 
\usepackage[top=2.8cm,bottom=2.8cm,left=2.cm,right=2.cm]{geometry}  
\usepackage{amsmath,amsfonts,amssymb}  

\usepackage{graphicx,subfig}  
\graphicspath{{./Figures/}{./PSTricks/}}  
\usepackage{multirow}
\usepackage{appendix}
\usepackage{wrapfig}
\usepackage{setspace}
\usepackage{cite}

\usepackage{csquotes}

\usepackage{array,booktabs}

\usepackage{enumitem}
\usepackage{caption3} 
\DeclareCaptionOption{parskip}[]{}
\usepackage[font=small,labelfont=bf]{caption}
\usepackage[final]{pdfpages}
\usepackage[colorinlistoftodos]{todonotes}
\usepackage{tikz}
\usepackage{sidecap}
\sidecaptionvpos{figure}{t}
\usepackage{abstract}

\newcommand{\red}{\color{red}}

\usepackage[pdftex]{hyperref} 

\title{\textbf{Space division multiplexing chip-to-chip quantum key distribution}} 
\author{Davide Bacco$^1$$^\dagger$, Yunhong Ding$^1$*, Kjeld Dalgaard$^1$, Karsten Rottwitt$^1$, and Leif Katsuo Oxenløwe$^1$}
\affil{$^1$ \small Department of Photonics Engineering, Technical University of Denmark, 2800 Kgs.~Lyngby, Denmark.}
\date{\vspace{-1em} \small   Dated: \today } 

\begin{document}

\pagestyle{plain}
\setcounter{page}{1}
\twocolumn[ 
\begin{@twocolumnfalse}
\maketitle
     \vspace{-0.8cm}
  \begin{abstract}
      \normalsize
         \vspace*{-1.0em}
\noindent Quantum cryptography is set to become a key technology for future secure communications. However, to get maximum benefit in  communication networks, transmission links will need to be shared among several quantum keys for several independent users. Such links will enable switching in quantum network nodes of the quantum keys to their respective destinations. In this paper we present an experimental demonstration of a photonic integrated silicon chip quantum key distribution protocols based on space division multiplexing (SDM), through multicore fiber technology. Parallel and independent quantum keys are obtained, which are useful in crypto-systems and future quantum network.

\end{abstract}
  \end{@twocolumnfalse}
 ]


\subsection*{Introduction}
In\blfootnote{$^\dagger$dabac@fotonik.dtu.dk, * yudin@fotonik.dtu.dk} contemporary society, communication security has become increasingly important. The security of the current cryptosystems, based on mathematical assumptions, are not guaranteed when quantum computers become available.  Quantum machines have already made indications that the current crypto codes can be easily eavesdropped~\cite{Shor1997}. This has spurred investigations into new security technologies based on quantum physics. In order to exchange secure information between users, Quantum key Distribution (QKD), a branch of Quantum Communications (QCs), provides good prospects for ultimate security based on the laws of quantum mechanics~\cite{BBPr,Scarani2009,Ma2005,Hwang2003,Bacco2016,Zhong2015}. 
In the last 30 years both free-space and fiber based QKD experiments, have demonstrated the exploitation of different physical principles. Furthermore, some cryptography companies are producing commercial devices that allow quantum security on a specific fiber link. However, most QKD systems are based on a point-to-point link, where the transmitter (Alice), and the receiver (Bob), generate a quantum key between two specific parties. In a future scenario, where QCs become standard technology, and where infrastructures, like banks and government buildings, will be connected through a quantum network, new principles in terms of key generation are required.
\begin{figure}
\vspace{0.2cm}
   \centering
    \includegraphics[width=8.5cm]{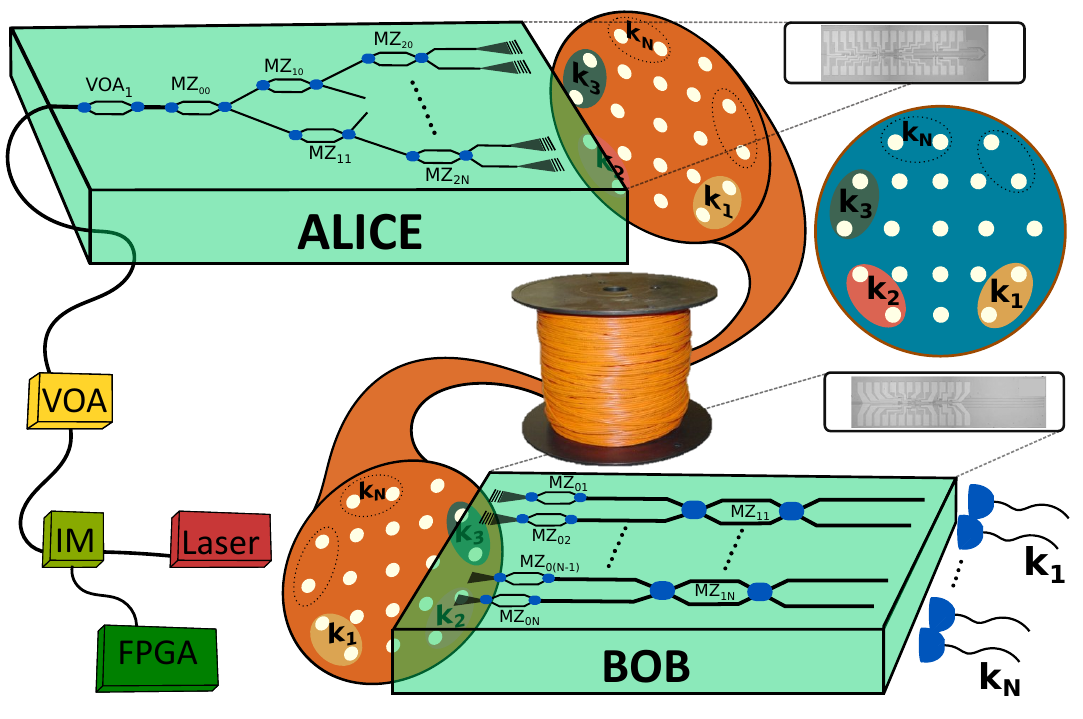}
    \caption{{\textit Scheme of the experiment}. A continuous wave (CW) laser beam at $1550$ nm is curved into pulses by an intensity modulator (IM) with a $5$ kHz repetition rate and a $10$ ns pulse width. The variable optical attenuator (VOA) decreases the mean number of photons per pulse ($\mu$) to be lower than one. On-chip Mach-Zehnder interferometers (MZIs), controlled by an FPGA, create the quantum states. A combination of VOA$_{1}$ and MZI$_{00}$ enable a decoy state technique with multiple and different numbers of photons per pulse. The quantum states are then measured independently by Bob's chip, creating an interference between pairs of cores. In this way parallel keys are generated between Alice and Bob. The rectangle insets show Alice's ($9.5 \times 2.5$ mm) and Bob's ($10 \times 2.5$ mm) silicon chips.}
    \label{fig:setup}
\end{figure}
The concept of a QKD network where customers need parallel independent keys, connecting multiple end-users and different nodes, will be highly useful. Here we describe a possible scenario for a quantum metropolitan/local area network (QMAN/QLAN), based on space division multiplexing (SDM) technique through a multicore fiber device.
Singular properties of quantum physics, like entangled photons, can be exploited for quantum teleportation and entanglement swapping protocols, with the purpose of key generation in a point to multi-point network~\cite{Zhu2015}. These networks, despite being very attractive from a security point of view, are problematic in terms of system requirements. Indeed, high rate entanglement sources, stable environments and a very low noise photon-detector are crucial for creating long distance links.
Alternatively, in the case of high-capacity demand, other principles can be adopted. Weak coherent pulses (WCP), using an attenuated laser with a mean number photon per pulse lower than one, is the most implemented technique being used in present days QKD systems. In these cases, several network schemes have been implemented and demonstrated. Usually, an active optical switch, is required in a point to multi-point implementation. 
Various switching dimension can be explored: wavelength division multiplexing (WDM), code division multiplexing (CDMA) and  time division multiplexing (TDM) are useful for implementation~\cite{Frohlich2013,Zhang2013,Dynes2016,Smania2016,Autebert2016,Hentschel2014,Nishioka2002}. However, all of these methods require extra devices on the line that introduces additional losses and cross-talk, which can compromise the final security.
Seen from this perspective, the possibility of using new low-cross talk technologies like multicore fibers (MCFs), well known in classical optical communications, is very promising. A recent work already demonstrated how a multicore fiber may be used in a communication link, to increase the secret key rate~\cite{Ding2016H}. Moreover, MCFs permit simultaneous transmission of classical and quantum channels with a very good signal-to-noise ratio (SNR) and isolation between cores, guaranteeing greater stability and robustness of the system, as well as allowing for strictly independent channels transmitted through the same fiber. 
In this paper we propose a solution for generating parallel and independent quantum keys using a silicon chip transmitter and exploiting the concept of space division multiplexing in a multicore fiber.  By adopting a single laser source and two different silicon chips, we realized a proof of concept (POC) experiment that demonstrates the generation of multiple independent quantum keys through the decoy-state BB84 protocol. This demonstrated functionality is a first step towards SDM quantum networks.

\section*{Experimental implementation}

\subsubsection*{QKD protocol}
The protocol implemented in the current experiment is the well known BB84 (with decoy states). By using the spatial dimension as a degree of freedom, instead of the standard way of using polarization, we encode the qubits on multiple cores of the MCF in such a way that for every two cores (cores $A$ and $B$ and cores $C$ and $D$ and so on), two mutually unbiased bases can be generated. In particular, for cores $A$ and $B$, the basis $\mathcal{X}_1$ is defined as ($\ket{A}; \ket{B}$) and basis $\mathcal{Z}_1$ as ($\ket{A+B}; \ket{A-B}$). Similarly for cores $C$ and $D$ the states $\lbrace \ket{C}, \ket{D} \rbrace \in \mathcal{X}_2$ and $\lbrace \ket{C+D}, \ket{C-D} \rbrace \in \mathcal{Z}_2$.
The final secret key rate is established using~\cite{Scarani2009}:
\begin{equation}
R \: \geq \: I_{AB}- min(I_{AE},I_{BE})
\label{eq:rate}
\end{equation}
$I_{AB}$ represents the classical mutual information between Alice and Bob ($I_{XY} = H(X)-H(X \vert Y)$), with the marginal entropy is defined as $H(X) = \sum_{x \in X} {p(x) \log p(x)}$. The right term of equation~\eqref{eq:rate} $\min$($I_{AE}$ and $I_{BE}$), is related to the quantum mutual information between Alice and Eve or Bob and Eve.
Note that using the same chip structure a slightly different implementation is possible, i.e. asymmetric BB84 with decoy-states~\cite{Tomamichel2012,Bacco2012}. This protocol relies on two mutually unbiased bases, but does not use an equal probability for all quantum states.
In other words, one of the two bases ($\mathcal{X}$), is chosen more often than the other ($\mathcal{Z}$) $p_\mathcal{X} \neq p_\mathcal{Z}$. In this way $\mathcal{X}$ is used for the key generation process and $\mathcal{Z}$ for security check. It follows that this protocol is more efficient compared to the standard BB84 (efficiency of $50 \%$), and it allows a higher final secret key rate. In the current experiment we selected an equal probability both for the bases choice and for the state preparation, so the overall efficiency.

\subsubsection*{Decoy-state weak coherent pulse generation}
Most practical QKD systems today are implemented with weak coherent pulses (WCP), generated by an attenuated laser. This scheme however, is not completely secure against particular kinds of attack, like photon-number splitting (PNS). In PNS attack, Eve blocks and discards all the single photon pulses while she only measures the multi-photon ones after the information reconciliation process. In this way, Bob and Eve measure the same quantum state, and at the end of the process Eve shares the same key. The decoy-state technique was introduced in order to overcome this problem. A controlled real-time fluctuation of the mean photon number per pulse ($\mu$) is used, in order to ensure the complete security of the final secret key. This technique is implemented in our experiment, where Alice's silicon chip, constituted by multiple Mach-Zehnder interferometers (MZIs), allows a complete freedom in terms of photon per pulse. By tuning the VOA$_1$ (variable optical attenuator) and the MZI$_{00}$ (the first index $0$ represents the level of the MZI starting from left, while the second index is related to the number of the cores of the fiber) with a specific voltage, different values of $\mu$ can be obtained (see Figure~\ref{fig:setup}). In table~\ref{tab:meanphoton} we reported all the different cases for a $2$-keys example. The MZI operates like a tunable ratio (transmittance/ reflectance) beam-splitter where Alice randomly decides which values to use.  In such a way, it is possible to create two independent quantum channels, which will generate two quantum keys. The example can be easily extended to a generic case where $N$ cores generate $N/2$ different keys.
\begin{table}[!h]
\centering
 \fontsize{12}{12}\selectfont 
\caption{Example of a two user system with the complete set of two decoy-levels and signal states ($u$, $v$ and a vacuum).}
\begin{tabular}{l c c c c c c c c r }
  \addlinespace[0.5ex]
  \hline			
  \addlinespace[0.5ex]
  {\bf key$_1$} & $u_1$ & $u_1$ & 0 & $v_1$ &  $u_1$ & $0$ & $0$ & $v_1$ & $v_1$\\
  \addlinespace[0.5ex]
  {\bf key$_2$} & $u_2$ & $0$ & $u_2$ & $u_2$ & $v_2$ & $0$ & $v_2$ & $0$ & $v_2$ \\
  \addlinespace[0.25ex]
  \hline  
\end{tabular}
\label{tab:meanphoton}
\end{table}

\subsubsection*{Generation of the quantum states}
The quantum states used in the current implementation are based on spatial encoding, exploiting different cores of a MCF. As shown in Figure~\ref{fig:setup}, a $1550$ nm continuous wave (CW) laser, has its light carved out to pulses by an intensity modulator at $5$ kHz repetition rate and pulse width of $10$ ns, which is coupled through a vertical coupler into the transmitter silicon chip (Alice). The quantum states are randomly prepared, by tuning the various MZIs, with a pseudorandom binary sequence (PRBS) sequence created by an FPGA board. Two PRBS seeds were used in order to create two parallel independent keys. In particular, by applying a different voltage on the MZI in Alice chip is possible to control the outputs of the integrated interferometers. After a first characterization of Alice's chip, we fixed a $0$ V level corresponding to having light only in one output (upper or lower). Consequently, a $V_{\pi}$ V value determines a reverse exit and a value of $V_{\pi}/2$ V represents the fifty-fifty case with light in both outputs.

Moreover, a real-time individual decoy states value is prepared for each pulse. Different voltages applied to the MZI$_{00}$ correspond to a specific decoy value, as reported in Table~\ref{tab:meanphoton}. Subsequently to the preparation of the  quantum states, we used a grating coupler array to couple from the silicon integrated circuit to a 7-cores fiber~\cite{Ding2015,VanLaere2009}. 
By exploiting this technique, we obtained a negligible cross talk between cores, around $-30$dB, and stable transmission can be achieved. The insertion and coupling losses attributed to Alice's chip are around $15$ dB. In this way, we created two independent quantum channels based on the principle of space division multiplexing.

\subsubsection*{Detection}
\begin{figure}[!ht]
\begin{center}
\includegraphics[width=8cm]{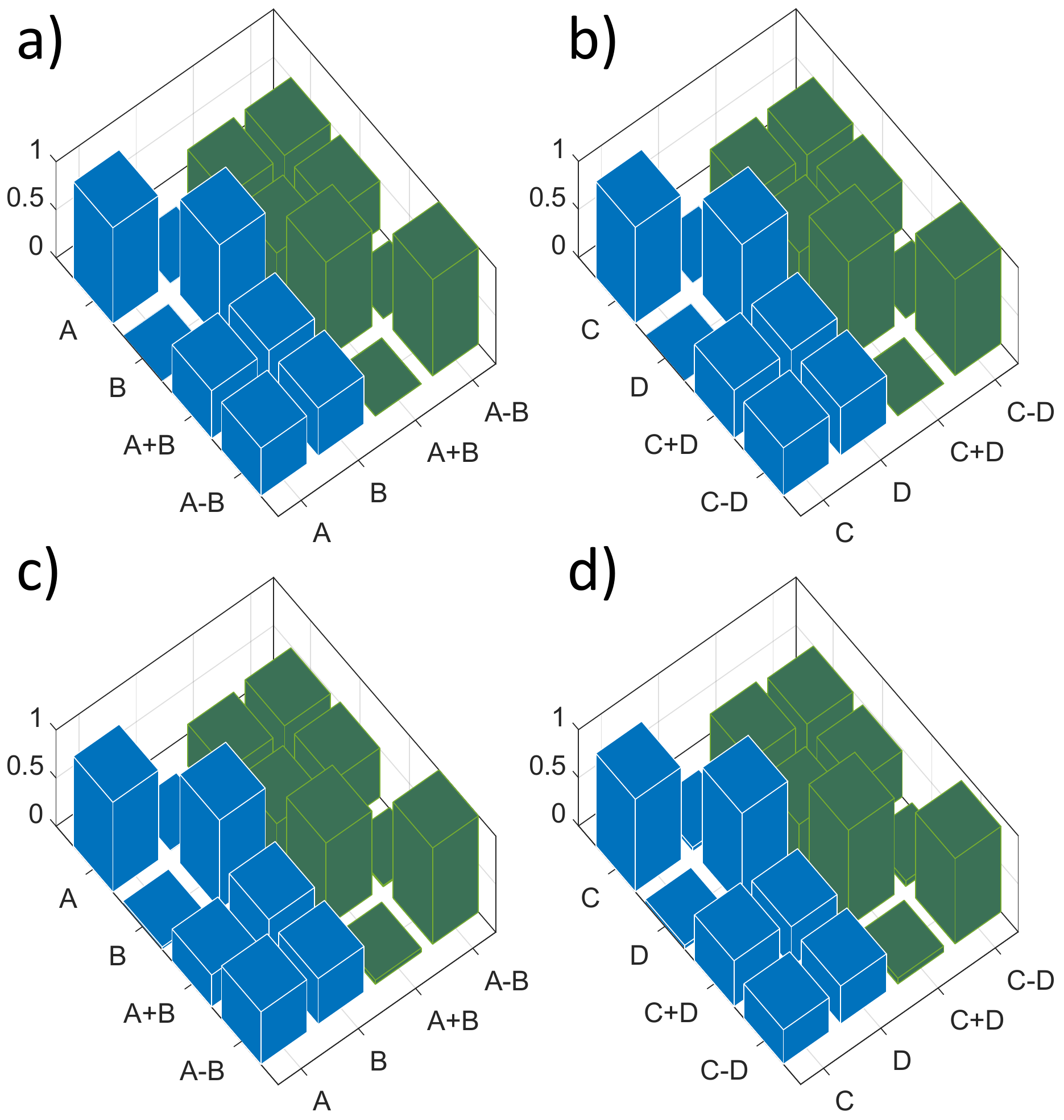}
\end{center}
\caption{\textit{Mutually unbiased bases characterization}. (a) Simulated MUBs for key number $1$; (b) Simulated MUBs for key number $2$; (c) Experimental data for key number $1$; (d) Experimental data for key number $2$. Each column corresponds to $30$ s of measurement with average $\mu$ of $0.4$ photon/pulse. Measured classical fidelity of $0.933$ for (c) matrix and $0.964$ for (d) matrix.}
\label{fig:tomography}
\end{figure}

\begin{figure}[!ht]
\begin{center}
\includegraphics[width=8cm]{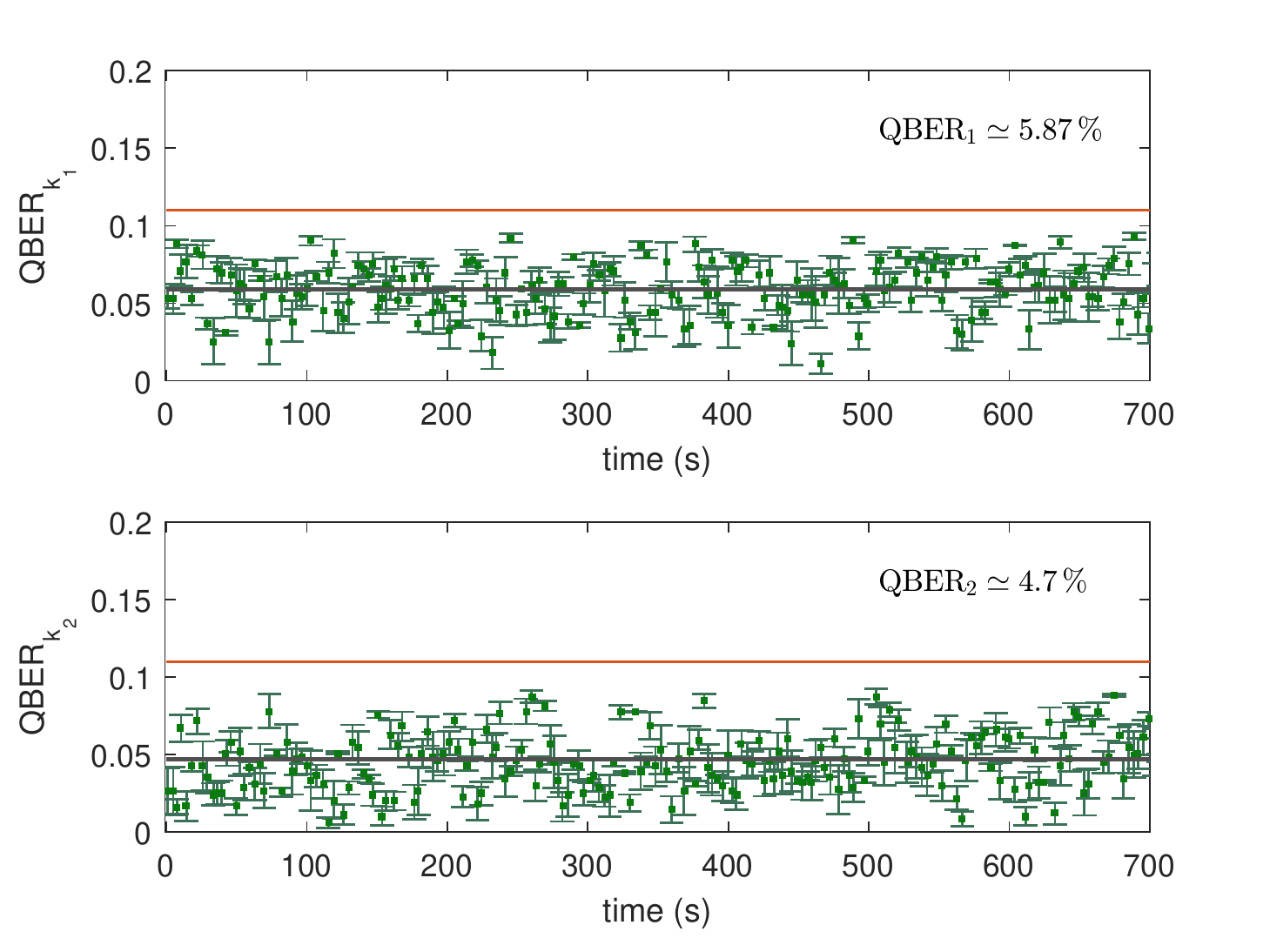}
\end{center}
\caption{\textit{Experimental bit error rate}. QBER  for $12$ minutes of acquired data for key $1$ and $2$. The gray lines represent the average QBER of the corresponding quantum keys ( $5.9 \, \% \pm 8.4 \cdot 10^{-3} $ and $4.7 \, \% \pm 8.8 \cdot 10^{-3}$ respectively). Orange line highlights the value of coherent attack limit in case of one-way reconciliation process ($11 \%$). Average $\mu_1$ and $\mu_2$ are $0.5 \pm 0.06 $ and $0.45 \pm 0.054 $ photon/pulse respectively.}
\label{fig:qqber}
\end{figure}

Once the quantum states are created and sent through the MCF, Bob measures the states in order to extract the quantum keys. In the experimental setup, two independent quantum keys, $k_1$ and $k_2$, as reported in Figure\ref{fig:setup}, are generated and the keys can be extracted by creating interference between the cores at the output~\cite{Ding2013,Ding2016}. In particular, tuning MZI$_{11}$ to MZI$_{1N}$, on Bob's side, it is possible to project the quantum states in different bases. Separate MZIs are used to measure in the mutually unbiased bases. In this way the randomness is maintained on the measurement side. The other MZIs (MZI$_{01}$ to MZI$_{0N}$), present on Bob's chip are used for phase stabilization between cores. In Figure~\ref{fig:tomography} we show the tomography of the two independent MUBs measured with weak laser pulses, repetition rate of $10$ kHz, and average mean photons number of $0.4$. By using the classical definition of fidelity ($F(x,y) = \sum_i (p_i q_i)^{1/2}$ with $x$ and $y$ random variables and $q_i$ and $p_i$ vectors of probability distribution) we measured $93$\% and $96$\%.
Another important parameter on the detection part is represented by the losses on the Bob's side. The insertion loss attributed to Bob's chip are measured to be around $8$ dB (from the output of the MCF, just before the facet, to the output of the chip). This loss can be further decreased in future chips by introducing an Al mirror below the grating area~\cite{Ding_mirror}. The four different outputs are coupled using a grating coupler array to four InGaAs single photon detectors, two ID230 and two ID220 respectively. In Figure~\ref{fig:qqber}
we report the measured QBER for the two independent keys. Stable and low QBER well below the coherent attack limit are obtained for more than $12$ minutes. The two plots show the different independent keys extracted in the experiment. 

\subsubsection*{Secret key rate}
\begin{figure}[!ht]
\begin{center}
\includegraphics[width=8cm]{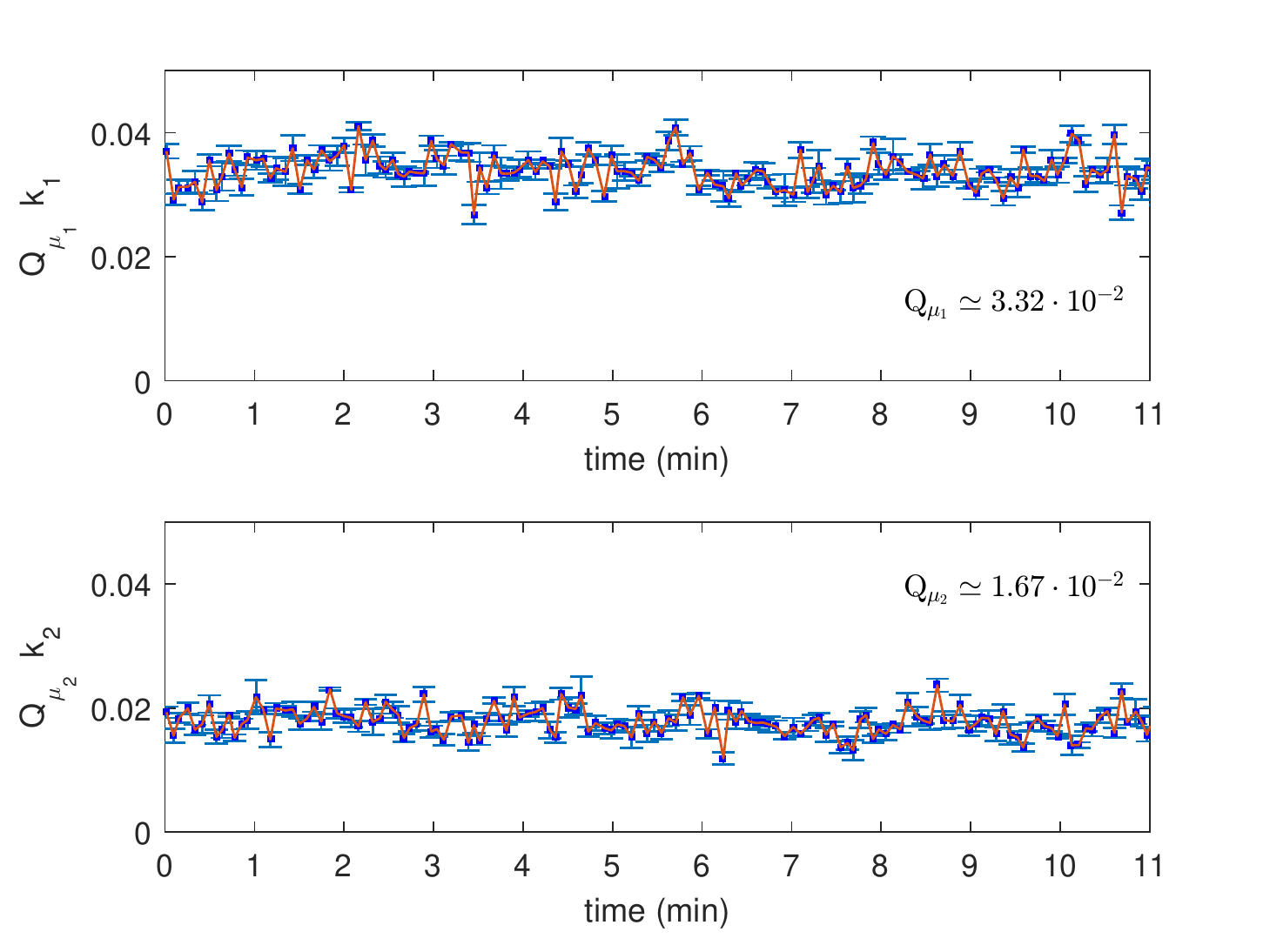}
\end{center}
\caption{\textit{Decoy state gain}. Gain of two independent decoy state keys for $11$ minutes of acquired data. Average gain values of $Q_{\mu_1} = 3.32 \cdot 10^{-2} \pm 1.2 \cdot 10^{-3} $ and $Q_{\mu_2}=1.67 \cdot 10^{-2} \pm 0.1 \cdot 10^{-3}$, $Q_{v_1}=1.8 \cdot 10^{-2} \pm 1 \cdot 10^{-3}$ and $Q_{v_2}=0.9 \cdot 10^{-2} \pm 9 \cdot 10^{-3}$.}
\label{fig:gain}
\end{figure}

After the measurement process it is possible to define a bound on the final secret key rate. This rate, given in Equation~\eqref{eq:rate}, depends on the strategy of the eavesdropper. We here included the case of collective attacks (CAs), where Eve can store the quantum states in her quantum memories and postpone the measurement till same future time. Alice and Bob discard the unmatched bases measurements, and subsequently perform error correction and privacy amplification, to extract the final key rate. In the case of decoy-state quantum key distribution it is possible to derive the following equation for the secret key rate:
\begin{equation}
R_{sk} \geq \tfrac{1}{2} \{- Q_u f(E_u) h_2(E_u) + Q_1 [1-h_2 (e_1)]  \}
\end{equation}
Here {\small$1/2$} is the probability related to the bases choice, $h_2$ is the binary Shannon information function, $u$ denotes the intensity of the signal states, $Q_u$ is the gain of the signal states, $E_u$ is the overall quantum bit error rate (QBER), $e_1$ is the error rate of the single-photon states and $f(x)$ is the bidirectional error correction efficiency, usually upper bounded with the value of $1.22$.
The parameter $Q_u$ and $E_u$ can be measured directly from the experiment, while $Q_1$ and $e_1$ can be estimated. Following the approach reported in Ma et al.~\cite{Ma2005}, it is possible to derive a secret key rate bound. To be noted that in a practical implementation of this system, a different bound including the statistical fluctuation can be used~\cite{Zhang2017}.
In Figure~\ref{fig:gain}, a real time measurement of the decoy state gain is reported.  An average value of $Q_{\mu_1} = 3.32 \cdot 10^{-2} \pm 1.2 \cdot 10^{-3} $ and $Q_{\mu_2}=1.67 \cdot 10^{-2} \pm 0.1 \cdot 10^{-3}$ are measured on Bob's side for the two independent keys, corresponding to a secret key rate generation of $113$ bit/s for $k_1$ and $60$ for $k_2$. Note that for a complete QKD system realization, where Eve cannot steal any information from the link, the gain value should be measured on Alice's side. However, in the current chip realization an extra output to do this measurement was not available. Nonetheless, in order to prove the real-time decoy state technique, we characterized the chip before the transmission over the multicore fiber channel in order to estimate the expected values. 

\subsection*{Discussion} 
MCFs represent the new frontier of optical communication which can be used for long distance high capacity transmission~\cite{Mizuno2016}. As previously proved in Ding et al.~\cite{Ding2016H}, this technology allows the creation of high-dimensional Hilbert space necessary in HD-QKD protocols. This PoC experiment extends the concept on a more general scenario. As seen from Figures~\ref{fig:tomography},~\ref{fig:qqber}, and~\ref{fig:gain}, the proposed scheme in principle works well for several minutes of measurement. In its present configuration, the experimental setup is merely for proof of principle, and higher key rates and longer transmission distances are expected to be achievable with minor upgrades. As a matter of fact, we fixed Alice's repetition rate to $5$ kHz. This choice was related to the transition time of the MZI. The interferometers are controlled by  heating, which is a very precise, stable and high contrast method (more than $30$ dB extinction ratio can be achieved), but comes with long rise and fall times. There exist several other solutions, based on modifying the material compositions and structures of the interferometers~\cite{Gan2015,Yan2016,Png2004}, which could be adapted to our scheme. InGaAs fast switches have recently been introduced and p-n junction can also be considered for silicon photonic devices. Once this technology gap will be resolved, the secret key rate and thereby the capacity of the QKD system will be improved ~\cite{Sibson2015,Ma2016}. 
Furthermore, our experiment was realized on an optical table with Alice and Bob being separated by only  a  few meters of multi-core fiber. This is by no means a limit, as already shown by Canas et al in~\cite{Canas2016}, where fiber-caused changes to phase and polarization is alleviated using a phase stabilization setup. 

In addition, by using space division multiplexing another advantage is achieved compared to the polarization encoding scheme. In particular, by using a polarization based decoy-state BB84 protocol over four cores of the multicore fiber, an higher final secret key rate can be achieved. However, problems like polarization instability (due to temperature and mechanical stress on the fiber) and polarization alignment (independent reference systems for each core) must be achieved during the communication process. In space division encoding the phase relation of the quantum states is slowly changing during the time, and a simple feedback loop will permit a stable long-distance QKD link~\cite{Govind2011}.
Moreover, one of the main problems in the deployment of quantum technologies is the compatibility between standard and quantum systems. In particular, optical communication through fiber links is subjected to various effects. The most critical one is represented by the Raman effect: inelastic scattering of photons by matter. In the case of high power monochromatic light propagating in an optical fibre, spontaneous Raman scattering transfers some of the photons to new frequencies. This problem, usually handled with narrow filters in classical optical communication, decreases the performance of the quantum systems by lowering the final key rate and the maximum distance. A solution is represented by the MCFs technology, where one of the cores (or more) can be used for classical light and the other ones as quantum channels~\cite{Dynes2016}.

\begin{figure}[!ht]
\begin{center}
\includegraphics[width=8cm]{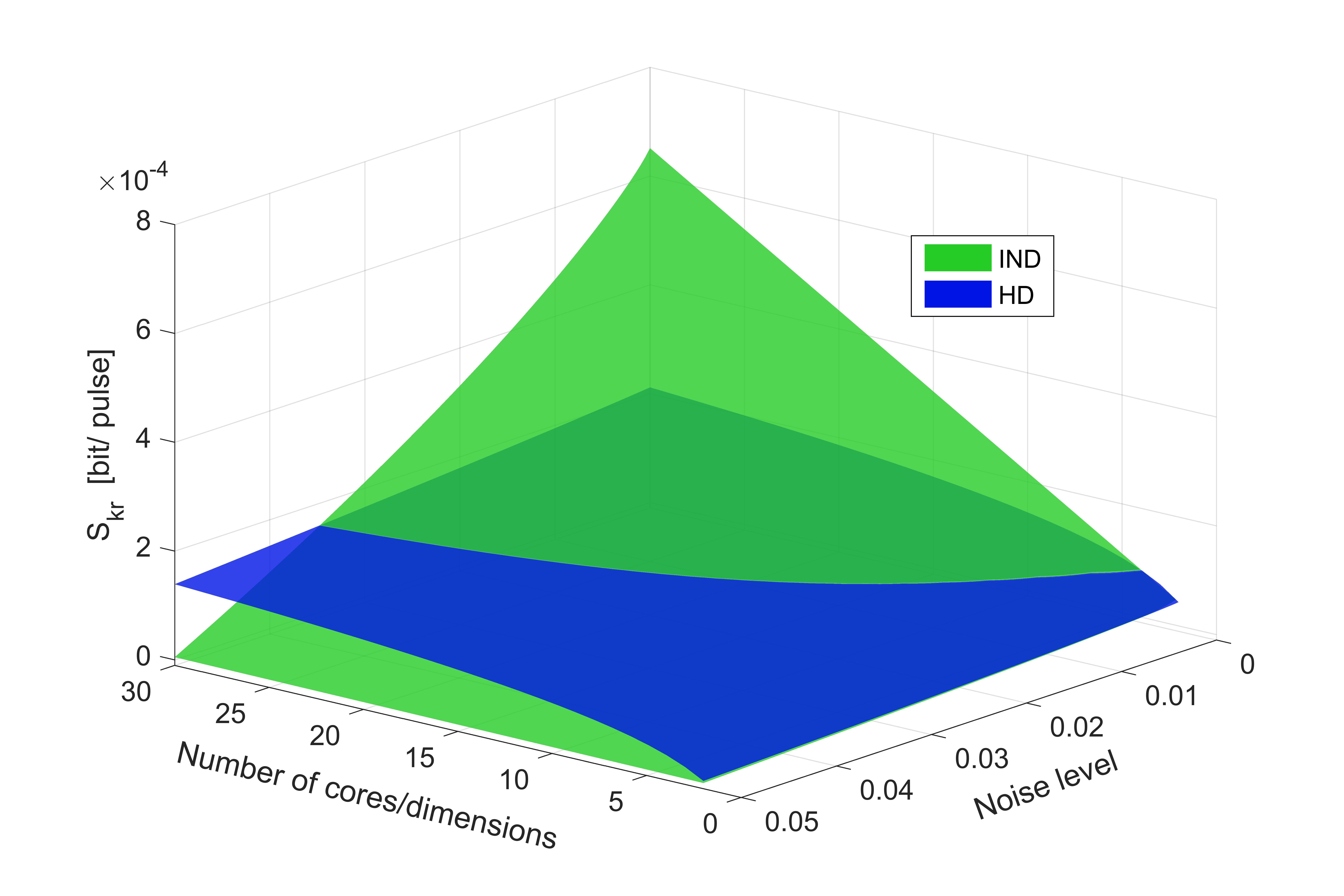}
\end{center}
\caption{\textit{Comparison between multiple independent quantum keys and HD-QKD}. Secret key rate as a function of the number of cores or dimensionality of the HD protocol, compared with a different intrinsic noise level for a fixed link distance of $50$ km. Simulation parameters $\eta_d=0.1$, $p_d=2 \cdot 10^{-8}$, $\alpha_{fiber}=0.37$ dB/km, $\eta_{BOB}=8$ dB.}
\label{fig:comparison}
\end{figure}

In this context we would point out that the presented scheme, based on spatial division multiplexed, can play an important role on future QKD systems. In fact, optical networks based on SDM are implemented and used in classical optical communication. HD-QKD based on SDM, like Higher Order Modes (HOM) and Orbital Angular Momentum (OAM) states, permits the creation of very high dimension Hilbert space. In case of HD system, the maximum acceptable QBER value depends on the quantity $N$, the dimension of the space (e.g. individual attack limit of $25\%$ for $N=4$ and $2$ MUBs). Regarding the final key rate, the number of bits that can be extracted scales with the equation $R \approx log_2 (N) [1-exp(-\eta )]$ (with $\eta =10^{-\alpha \cdot l/10} $ and $l$ the link distance). In the case of SDM instead, the key rate is linearly dependent with the number of cores involved, $R \approx (N/2) [1-exp( -\eta )]$. Furthermore, a comparison of the achievable rate, obtained with different multiplexing techniques (WDM, TDM, CDMA), must be introduced. In the case of WDM setup, the final rate can be approximated to $R \approx (N) [1-exp( -\eta )]$ where $N$ in this case represents the different wavelengths used in the system (assuming perfect filters). To be noted, as already explained, that a WDM system requires $N$ different transmitters, and very good filter in the receiver side, in order to avoid cross-talk between adjacent channels. The final rate for TMD can be written as $R \approx (N) [1-exp( -\eta /N )]$, assuming a $N$ users and perfect splitter. Finally, considering the case of CDMA technique (optimal orthogonal codes), the key rate can be reported as $R \approx [(1-w^2)/N_c]^{N-1} [1-exp( -\eta /N )]$ with $w$ defined as the weight of the implemented code and $N_c$ the length of the code~\cite{Razavi2011}.
As a consequence, and as reported in Figure~\ref{fig:comparison}, depending on the quality of the channel and on the setting in which the system is used (distance, noise, temperature instability, etc..), a choice between the HD-QKD and independent quantum keys is expected.

\subsection*{Conclusion}
In this paper, we proposed and demonstrated the use of multicore fibers used with SDM technique for QKD transmission. We successfully proved the principle by sending two separate quantum keys prepared by a silicon photonic chip through the same multicore fiber, and receiving the keys through a second silicon photonic chip. The measured QBER confirms the correct transmission and interpretation of the QKD scheme.

\subsection*{Methods}
\textbf{Device realization}
Alice and Bob photonic integrated circuit (PICs) are formed by $250$ nm of SOI silicon thickness and $1$ $\mu$m of buried oxide layer (BOX) We used a single step process of e-beam lithography and inductively coupled plasma (ICP). Subsequently $1.500$ $\mu$m thick layer of SiO2 was deposited on top of the chip using plasma-enhanced chemical vapor deposition technique. 
After the polish process the layer of SiO2 was reduced to $1$ $\mu$. In this way SiO2 works as a isolation layer between the silicon waveguide and the Titanium heaters fabricated on a second time to avoid potential optical losses. In this way by using e-beam lithography followed by metal deposition and liftoff process we created $100$ nm thick of titanium heaters. As last step UV lithography technique, followed by metal deposition and liftoff process, was used to fabricate Au/Ti contact layer. The chip was then cleaved and wire-bonded to a PCB board.

\textbf{Electronic design}
The chip-to-chip parallel key QKD scheme, based on space-division multiplexing is feasible thank to a real time control of the different MZIs presented on the silicon chip. These MZIs, as reported above, are controlled by heaters: conductor material which change his property when a voltage is applied. In order to tune in real-time these MZIs, different electrical signals are required in the transmitter and receiver side. An Altera FPGA board emits $8$ digital parallel outputs every $0.2$ ms, which are converted into analog voltages by $8$ digital-analog converters (DACs). Then, these analog signals are send to the transmitter and the receiver PCB board by flat cables.  

\subsection*{Acknowledgements}
This work is supported by the Danish Council for Independent Research (DFF-1337-00152 and DFF-1335-00771), by the Center of Excellence, SPOC (Silicon Photonics for Optical Communications (ref DNRF123) and from the People Programme (Marie Curie Actions) of the European Union's Seventh Framework Programme (FP7/2007-2013) under REA grant agreement n$^\circ$ $609405$ (COFUNDPostdocDTU).

\subsection*{Author contributions}
D. Bacco and Y. Ding proposed the idea. Y. Ding designed and fabricated the silicon PICs. K. Dalgaard and D. Bacco designed the electrical controlling circuits. D. Bacco, Y. Ding, and K. Dalgaard performed the system experiment. D. Bacco carried out the theoretical analysis on the proposed protocol. D. Bacco, Y. Ding, K. Rottwitt, and L. K. Oxenløwe discussed the results. All authors contributed to the writing of the manuscript.

\subsection*{Competing financial interests}
The authors declare no competing financial interests.

\titleformat*{\section}{\bf\normalsize}

\end{document}